\documentclass[12pt]{article}
\usepackage{jhep-mod}
\usepackage{bm}
\usepackage{amssymb,amsmath,amsthm}
\usepackage{mathrsfs}
\usepackage[utf8]{inputenc}
\usepackage{enumerate}
\usepackage{appendix}
\usepackage{graphicx}
\usepackage{dcolumn} 
\usepackage{bm}
\usepackage{multirow}
\usepackage{float}
\usepackage{tikz}
\usepackage{setspace}
\definecolor{purple}{rgb}{1,0,1}
\definecolor{lime}{HTML}{A6CE39} % needs xcolor
\definecolor{lime}{HTML}{A6CE39}
\newcommand{\orcidicon}{%
	\begin{tikzpicture}
	\draw[lime, fill=lime] (0,0) 
		circle [radius=0.16] 
		node[white] {{\fontfamily{qag}\selectfont \tiny ID}};
	\draw[white, fill=white] (-0.0625,0.095) 
		circle [radius=0.007];
	\end{tikzpicture}
	\hspace{-5mm}
}
\newcommand\orcidAlex{{\href{https://orcid.org/0000-0002-1763-3563}{\orcidicon}}}
\newcommand\orcidMatt{{\href{https://orcid.org/0000-0003-1088-6485}{\orcidicon}}}
\begin{document}
%========================================================
\title{\huge{Black-bounce to traversable wormhole}}
%========================================================
\author{\Large Alex Simpson\orcidAlex{}{\sf and} Matt Visser\orcidMatt{} }
%%%%%%%%%%%%%%%%%%%%%%%%%%%%%%%%%%%%%%
\affiliation{School of Mathematics and Statistics, Victoria University of Wellington, \\ \null\quad PO Box 600, Wellington 6140, New Zealand}
%%%%%%%%%%%%%%%%%%%%%%%%%%%%%%%%%%%%%%
\emailAdd{alex.simpson@sms.vuw.ac.nz, matt.visser@sms.vuw.ac.nz}
%%%%%%%%%%%%%%%%%%%%%%%%%%%%%%%%%%%%%
\abstract{
\parindent0pt
\parskip7pt

So-called ``regular black holes'' are a topic currently of considerable interest in the general relativity and astrophysics communities. 
Herein we investigate a particularly interesting regular black hole spacetime described by the line element
\[
ds^{2}=-\left(1-\frac{2m}{\sqrt{r^{2}+a^{2}}}\right)dt^{2}+\frac{dr^{2}}{1-\frac{2m}{\sqrt{r^{2}+a^{2}}}}
+\left(r^{2}+a^{2}\right)\left(d\theta^{2}+\sin^{2}\theta \;d\phi^{2}\right).
\]
This spacetime neatly interpolates between the standard Schwarzschild black hole and the Morris--Thorne traversable wormhole;
at intermediate stages passing through a black-bounce (into a future incarnation of the universe), an extremal null-bounce (into a future incarnation of the universe), and a traversable wormhole.  As long as the parameter $a$ is non-zero the geometry is everywhere regular, so one has a somewhat unusual form of ``regular black hole'', where the ``origin'' $r=0$ can be either spacelike, null, or timelike. Thus this spacetime generalizes and broadens the class of ``regular black holes'' beyond those usually considered.

\medskip
{\sc Date:} 18 December 2018; 1 February 2019; \LaTeX-ed \today

%\medskip
%{\sc arXiv:} 1805.nnnnn

Morris--Thorne wormhole; regular black hole.

\medskip
{\sc Pacs}: 04.20.-q; 04.20.Gz; 04.70.-s; 04.70.Bw.
}
%=====================================================
\maketitle
%=====================================================
\def\d{{\mathrm{d}}}
\def\tr{{\mathrm{tr}}}
\parindent0pt
\parskip7pt

%=====================================================
\section{Introduction}
%====================================================

Ever since Bardeen initially proposed the concept of a regular black hole in 1968~\cite{Bardeen:1968}, see also the more recent references~\cite{Bergmann-Roman,Hayward:2005, Bardeen:2014, Frolov:2014, Frolov:2014b, Frolov:2016, Frolov:2017, Frolov:2018,Cano:2018,Bardeen:2018, regular, beyond}, the notion has been intuitively attractive due to its non-singular nature. When exploring various candidates for regular black hole geometries within the framework of general relativity, it pays to compile examples of various metrics of interest, and provide thorough analyses of their phenomenological properties~\cite{regular, beyond}. As such, we propose the following candidate regular black hole specified by the spacetime metric:
\begin{equation}
    ds^{2}=-\left(1-\frac{2m}{\sqrt{r^{2}+a^{2}}}\right)dt^{2}+\frac{dr^{2}}{1-\frac{2m}{\sqrt{r^{2}+a^{2}}}}+\left(r^{2}+a^{2}\right)\left(d\theta^{2}+\sin^{2}\theta d\phi^{2}\right).
\end{equation}
This spacetime is carefully designed to be a minimalist modification of the ordinary Schwarzschild spacetime.
Adjusting the parameter $a$, this metric represents either: 
\begin{enumerate}
\itemsep-3pt
\item 
The ordinary Schwarzschild spacetime; 
\item
A regular black hole geometry with a one-way spacelike throat;
\item
A one-way wormhole geometry with an extremal null throat; \\
(compare particularly with reference~\cite{Cano:2018}); or
\item
A canonical traversable wormhole geometry, \\
(in the Morris--Thorne sense~\cite{Morris:1988a, Morris:1988b, Visser:1989a, Visser:1989b, Lorentzian, Visser:2003, Hochberg:1997, Poisson:1995, Barcelo:2000, Hochberg:1998, Cramer:1994, Visser:1997, Barcelo:1999, Garcia:2011, Boonserm:2018, Lobo:2004}), with a two-way timelike throat.
\end{enumerate} In the region where the geometry represents a regular black hole the geometry is unusual in that it describes a bounce into a future incarnation of the universe, rather than a bounce back into our own universe~\cite{Barcelo:2014, Barcelo:2014b, Barcelo:2015, Barcelo:2016, Garay:2017, Rovelli:2014, Haggard:2015, Christodoulou:2016, DeLorenzo:2015, Malafarina:2017, Olmedo:2017, Barrau:2018, Malafarina:2018}. 
Conducting a standard analysis of this metric within the context of general relativity we find the locations of photon spheres and ISCOs for each case, calculate Regge--Wheeler potentials, and draw conclusions concerning the nature of the curvature of this spacetime.

%=====================================================
\section{Metric analysis and Carter--Penrose diagrams}\label{sec:metric}
%=====================================================
Consider the metric:
\begin{equation}\label{RBHmetric}
    ds^{2}=-\left(1-\frac{2m}{\sqrt{r^{2}+a^{2}}}\right)dt^{2}+\frac{dr^{2}}{1-\frac{2m}{\sqrt{r^{2}+a^{2}}}}+\left(r^{2}+a^{2}\right)\left(d\theta^{2}+\sin^{2}\theta d\phi^{2}\right).
\end{equation}
Note that if $a=0$ then this is simply the Schwarzschild solution, so enforcing $a\neq 0$ is a sensible starting condition if we are to conduct an analysis concerning either regular black holes or traversable wormholes (trivially,  the Schwarz\-schild solution models a geometry which is neither). Furthermore, this spacetime geometry is manifestly static and spherically symmetric.  That is,  it admits a global, non-vanishing, timelike Killing vector field that is hypersurface orthogonal, and there are no off-diagonal components of the matrix representation of the metric tensor; fixed $r$ coordinate locations in the spacetime correspond to spherical surfaces. This metric does not correspond to a traditional regular black hole such as the Bardeen, Bergmann--Roman, Frolov, or Hayward geometries \cite{Bardeen:1968, Bergmann-Roman, Hayward:2005, Bardeen:2014, Frolov:2014, Frolov:2014b, Frolov:2016, Frolov:2017, Frolov:2018, Cano:2018,Bardeen:2018}. Instead, depending on the value of the parameter $a$, it is either a regular black hole (bouncing into a future incarnation of the universe) or a traversable wormhole.

Before proceeding any further, note that the coordinates have natural domains:
\begin{equation}
r\in(-\infty,+\infty);\qquad 
t\in(-\infty,+\infty);\qquad
\theta\in [0,\pi];\qquad
\phi\in(-\pi,\pi].
\end{equation}
Analysis of the radial null curves in this metric yields (setting $ds^{2}=0$, $d\theta=d\phi=0$):
\begin{eqnarray}
    \frac{dr}{dt} = \pm\left(1-\frac{2m}{\sqrt{r^{2}+a^{2}}}\right) \ .
\end{eqnarray}
It is worth noting that this defines a ``coordinate speed of light" for the metric (\ref{RBHmetric}),
\begin{equation}
    c(r)=\left\vert\frac{dr}{dt}\right\vert=\left(1-\frac{2m}{\sqrt{r^{2}+a^{2}}}\right) \ ,
\end{equation}
and hence an effective refractive index of:
\begin{equation}
    n(r)=\frac{1}{\left(1-\frac{2m}{\sqrt{r^{2}+a^{2}}}\right)} \ .
\end{equation}
Let us now examine the coordinate location(s) of horizon(s) in this geometry:
\begin{itemize}
 
    \item If $a>2m$, then $\forall \ r\in(-\infty,+\infty)$ we have $\frac{dr}{dt}\neq 0$, so this geometry is in fact a (two-way) traversable wormhole~\cite{Morris:1988a, Morris:1988b, Visser:1989a, Visser:1989b, Lorentzian, Visser:2003, Hochberg:1997, Poisson:1995, Barcelo:2000, Hochberg:1998, Cramer:1994, Visser:1997, Barcelo:1999, Garcia:2011, Boonserm:2018, Lobo:2004}.
      
    \item If $a=2m$, then as $r\rightarrow 0$ from either above or below, we have $\frac{dr}{dt}\rightarrow 0$. Hence we have a horizon at coordinate location $r=0$. However, this geometry is not a black hole. Rather, it is  a one-way wormhole with an extremal null throat at $r=0$.

    \item If $a<2m$, then consider the two locations $r_\pm=\pm\sqrt{(2m)^{2}-a^{2}}$; this happens when $\sqrt{r_\pm^{2}+a^{2}}=2m$.  
    Thence
    \begin{eqnarray}
        \exists \ r_\pm \in\mathbb{R}: \qquad \frac{dr}{dt}=0.
    \end{eqnarray}
    That is, when $a<2m$ there will be symmetrically placed $r$-coordinate values $r_\pm = \pm|r_\pm|$  which correspond to a pair of horizons.
\end{itemize}
The coordinate location $r=0$ maximises both the non-zero curvature tensor components (see section \ref{sec:curvature}) and the curvature invariants (see section \ref{sec:invariants}). 

We may therefore conclude that in the case where $a>2m$ the traversable wormhole throat is a timelike hypersurface located at $r=0$, and negative $r$-values correspond to the universe on the other side of the geometry from the perspective of an observer in our own universe. 

\begin{figure}[!htb]
%\vspace{-1cm}
\begin{center}
\includegraphics[scale=0.50]{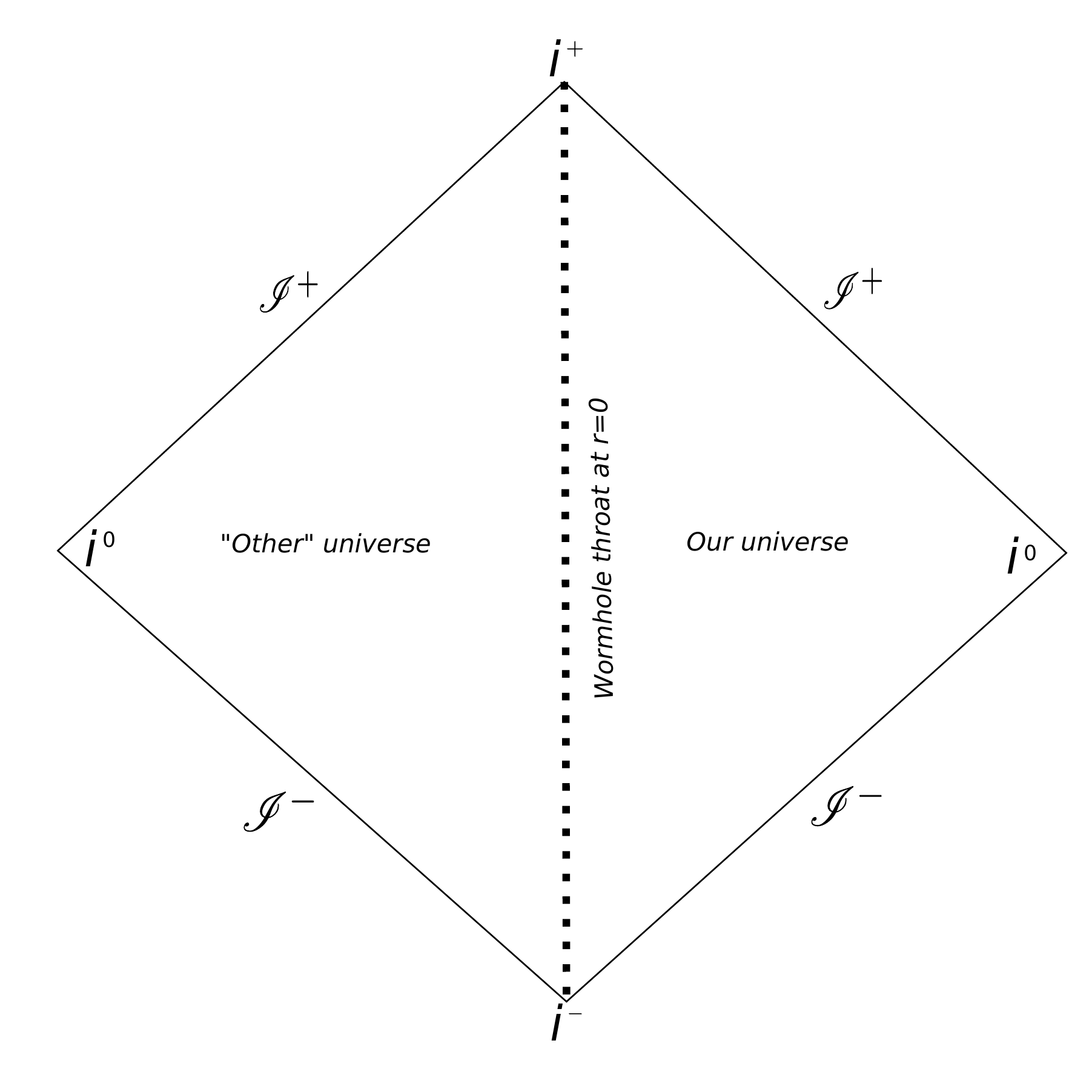}\qquad
\end{center}
{\caption{{Carter--Penrose diagram for the case when $a>2m$ and we have a traditional traversable wormhole in the Morris--Thorne sense}.}
\label{F:wormhole}}
\end{figure}

We then have the standard Carter--Penrose diagram for traversable wormholes as presented in figure~\ref{F:wormhole}.

\begin{figure}[!htb]
\vspace{-0.5cm}
\begin{center}
\includegraphics[scale=0.50]{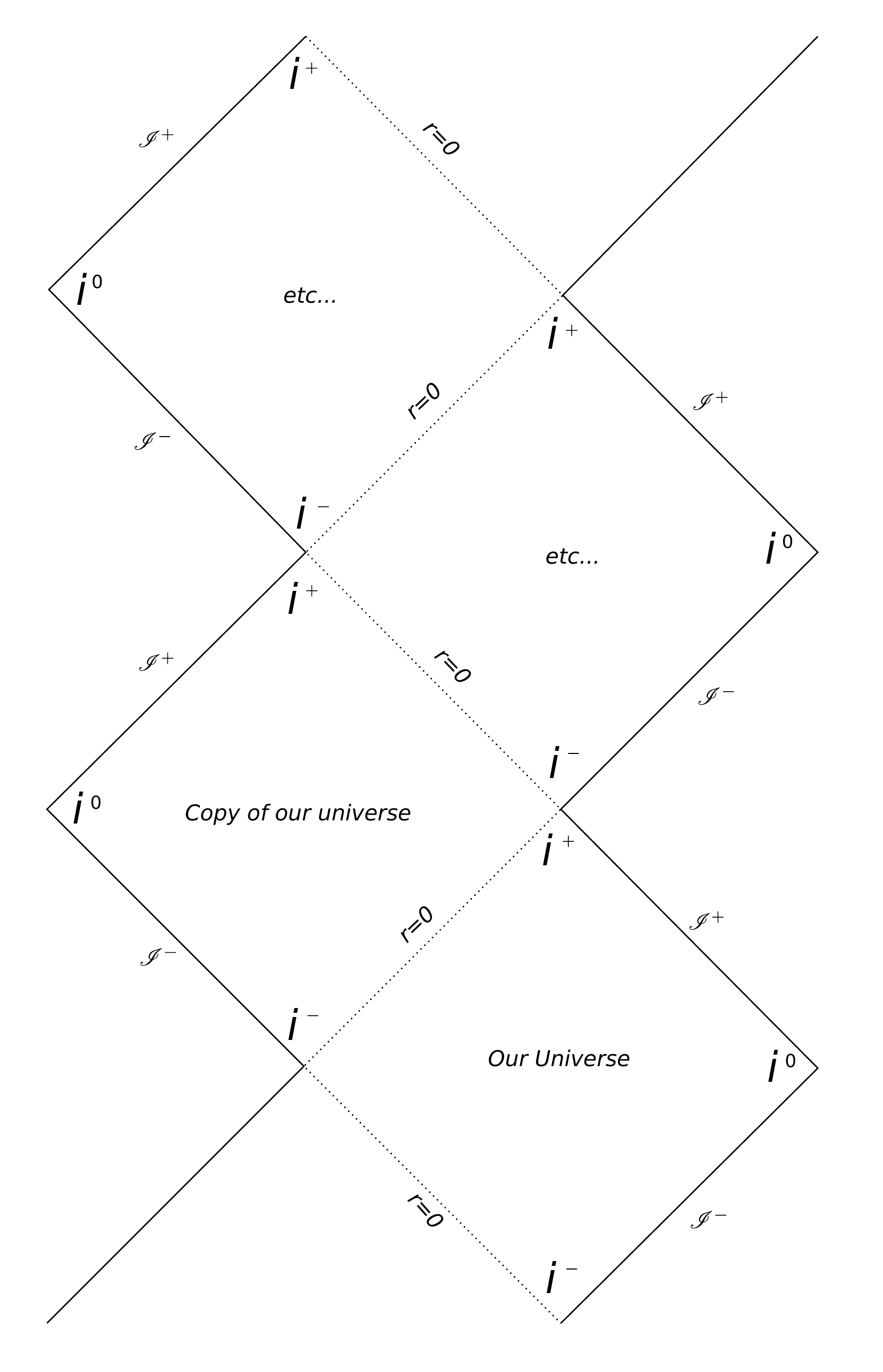}\qquad
\end{center}
{\caption{{Carter--Penrose diagram for the maximally extended spacetime in the case when $a=2m$. In this example we have a one-way wormhole geometry with a null throat}.}
\label{F:null-bounce-1}}
\end{figure}

Similarly for the null case $a=2m$ the null throat is located at the horizon $r=0$. Note that in this instance the wormhole geometry is only one-way traversable. 
The Carter--Penrose diagram for the maximally extended spacetime in this case is given in figure \ref{F:null-bounce-1}.

%=====================================================
%=====================================================

%=====================================================
\null
%=====================================================

\begin{figure}[!htb]
\begin{center}
\includegraphics[scale=0.50]{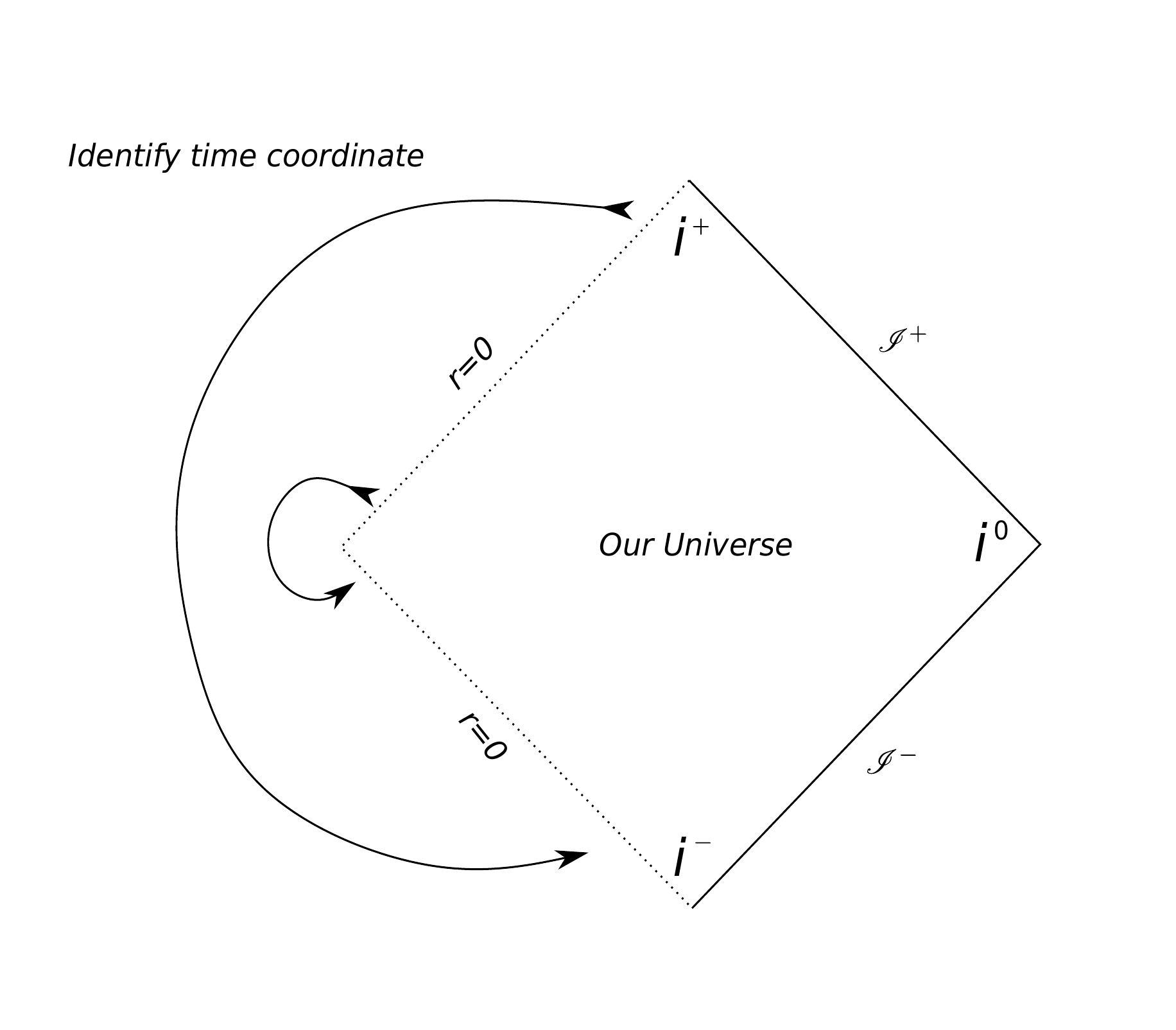}\qquad
\end{center}
{\caption{{Carter--Penrose diagram for the case when $a=2m$ where we have identified the future null bounce at $r=0$ with the past null bounce at $r=0$}.}
\label{F:null-bounce-2}}
\end{figure}

As an alternative construction we can identify the past null bounce at $r=0$ with the future null bounce at $r=0$ 
yielding the `looped' Carter--Penrose diagram of figure \ref{F:null-bounce-2}.

%=====================================================
%=====================================================

\begin{figure}[!htb]
\begin{center}
\includegraphics[scale=0.50]{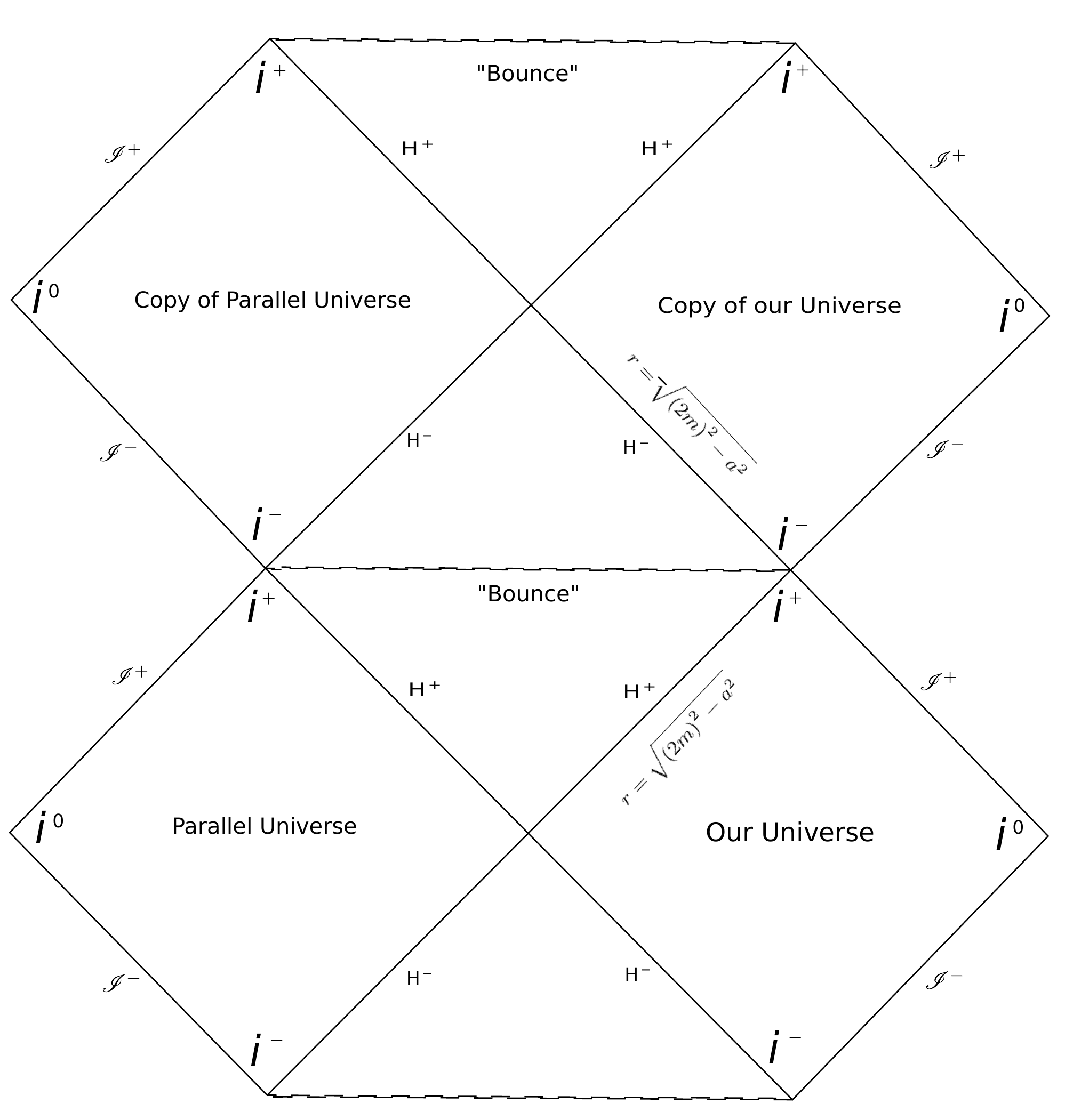}\qquad
\end{center}
{\caption{{Carter--Penrose diagram for the maximally extended spacetime when $a\in(0,2m)$. In this example the time coordinate
runs up the page, 
`bouncing' through the $r=0$  hypersurface in each black hole region into a future copy of our own universe \emph{ad infinitum}.}}\label{F:bounce-1}}
\end{figure}

For regular black holes, 
we can restrict our attention to the interval $a\in(0,2m)$. Then the hypersurface $r=0$ is a spacelike spherical surface which marks the boundary between our universe and a bounce into a separate copy of our own universe. For negative values of $r$ we have `bounced' into another universe.
See figure \ref{F:bounce-1} for the  relevant Carter--Penrose diagram.
(Contrast these Carter--Penrose diagrams with the standard one for the maximally extended Kruskal--Szekeres version of Schwarzschild --- see for instance references~\cite{Wald, MTW, Hell} --- the major difference is that the singularity has been replaced by a spacelike hypersurface representing a ``bounce''.)

%=====================================================
%=====================================================

\begin{figure}[!htb]
\begin{center}
\includegraphics[scale=0.50]{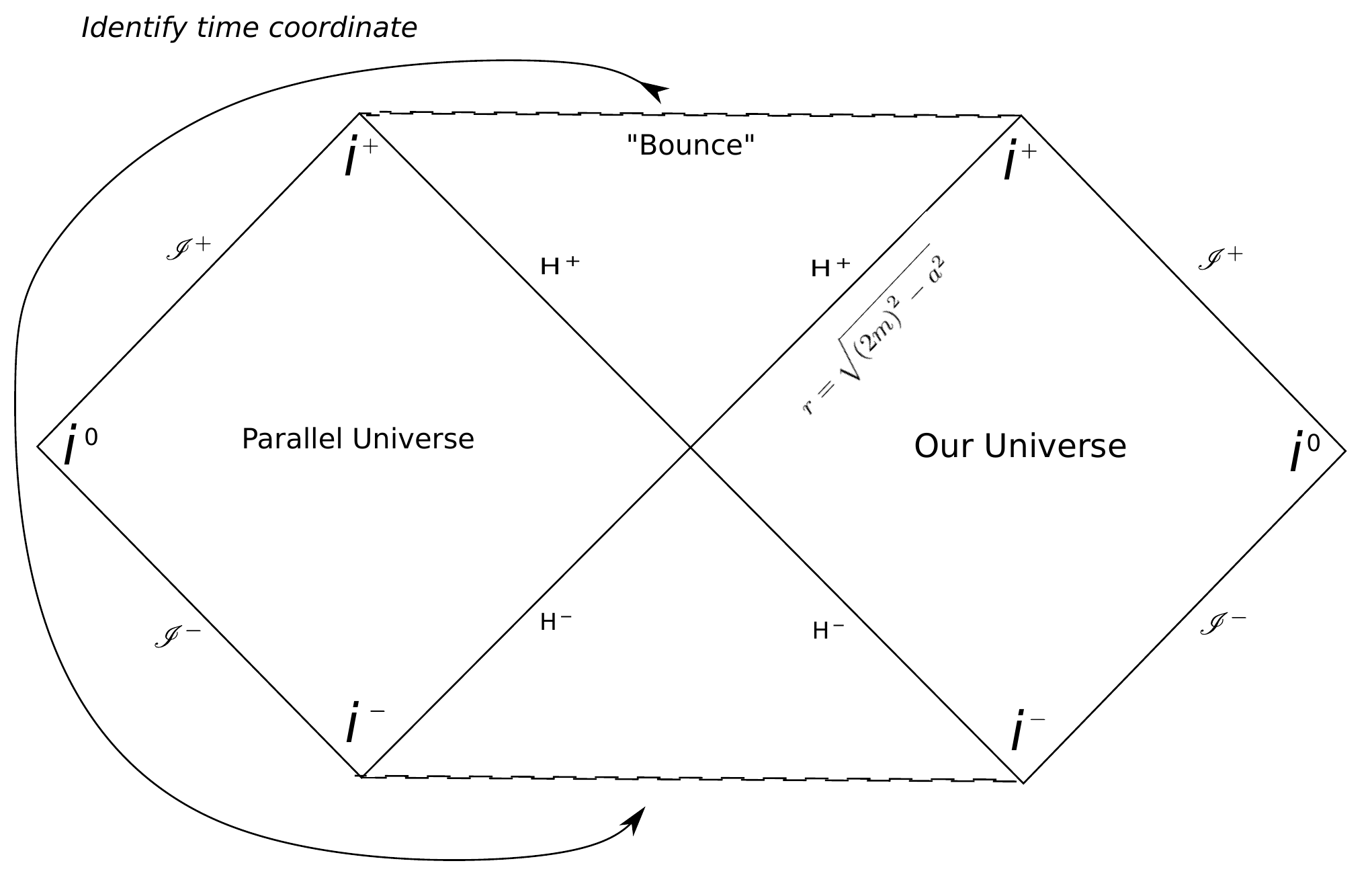}\qquad
\end{center}
{\caption{{Periodic boundary conditions in time when $a\in(0,2m)$. In this example we impose periodic boundary conditions on the time coordinate such that the future bounce is identified with the past bounce}.}
\label{F:bounce-2}}
\end{figure}

Another possibility of interest for when $a\in(0,2m)$ arises when the  $r=0$ coordinate for the `future bounce' is identified with the  $r=0$ coordinate for the `past bounce'. That is, there is still a distinct time orientation but we impose periodic boundary conditions on the time coordinate such that time is cyclical. This case yields the Carter-Penrose diagram of figure \ref{F:bounce-2}. (Note that the global causal structure is much milder than that for the so-called ``twisted'' black holes~\cite{twisted}.)

%=====================================================
%=====================================================

%====================================================
\section{Curvature tensors}\label{sec:curvature}
%====================================================

Next it is prudent to check that there are no singularities in the geometry, otherwise we do not satisfy the requirements for the regularity of our black hole. 
In view of the diagonal metric environment of (\ref{RBHmetric}) we can clearly see that the chosen coordinate basis is orthogonal though not orthonormal, and it therefore follows that the (mixed) non-zero components of the Riemann tensor shall be the same with respect to this basis as to any orthonormal tetrad, ensuring that the appearance (or lack thereof) of any singularities is not simply a coordinate artefact. 

With this in mind, for simplicity we first consider the non-zero components of the Weyl tensor (these are equivalent to orthonormal components):
\begin{eqnarray}
    C^{t\theta}{}_{t\theta} = C^{t\phi}{}_{t\phi} = C^{r\theta}{}_{r\theta} = C^{r\phi}{}_{r\phi} &=&  -\frac{1}{2}C^{tr}{}_{tr} = -\frac{1}{2}C^{\theta\phi}{}_{\theta\phi} \nonumber \\
    &=&  \frac{6r^2m+a^2\left(2\sqrt{r^2+a^2}-3m\right)}{6\left(r^2+a^2\right)^{\frac{5}{2}}} \ .
\end{eqnarray}
Note that as $r\to0$ these Weyl tensor components approach the finite value ${2a-3m\over 6 a^3}$. 

For the Riemann tensor the non-zero components are a little more complicated:
\begin{eqnarray}
    R^{tr}{}_{tr} &=& \frac{m(2r^{2}-a^{2})}{(r^{2}+a^{2})^{\frac{5}{2}}} ;
    \nonumber\\    
    R^{t\theta}{}_{t\theta} &=& R^{t\phi}{}_{t\phi}=\frac{-r^{2}m}{(r^{2}+a^{2})^{\frac{5}{2}}}; 
    \nonumber \\
    R^{r\theta}{}_{r\theta} &=& R^{r\phi}{}_{r\phi}=\frac{m\left(2a^{2}-r^{2}\right)-a^{2}\sqrt{r^{2}+a^{2}}}{(r^{2}+a^{2})^{\frac{5}{2}}}; 
    \nonumber \\
    &&\nonumber \\
    R^{\theta\phi}{}_{\theta\phi} &=& \frac{2r^{2}m+a^{2}\sqrt{r^{2}+a^{2}}}{(r^{2}+a^{2})^{\frac{5}{2}}}.
\end{eqnarray}

Provided $a\neq0$, as $\vert r\vert\rightarrow 0$ all of these Riemann tensor components approach finite limits:
\begin{eqnarray}
    && R^{tr}{}_{tr} \to -\frac{m}{a^3} \ ; \nonumber\\
    && R^{t\theta}{}_{t\theta} = R^{t\phi}{}_{t\phi} \to 0 \ ; \nonumber \\
    && R^{r\theta}{}_{r\theta} = R^{r\phi}{}_{r\phi} \to \frac{2m-a}{a^3} \ ; \nonumber\\
    &&  R^{\theta\phi}{}_{\theta\phi} \to \frac{1}{a^2} \ .
\end{eqnarray}
As $\vert r\vert$ increases, with $m$ and $a$ held fixed,  all components asymptote to multiples of $m/r^3$, hence as $\vert r\vert\rightarrow+\infty$, all components tend to $0$ (this is synonymous with the fact that for large $\vert r\vert$ this geometry models weak field general relativity).
Hence $\forall \ r\in(-\infty,+\infty)$ the components of the Riemann tensor are strictly finite. We may conclude that on the interval $a\in(0,2m]$ there is a horizon, but no singularity, and the metric really does represent the geometry of a regular black hole. In the case when $a>2m$ and we have a traversable wormhole, trivially there are also no singularities.

%====================================================

The Ricci tensor has non-zero (mixed) components:
\begin{eqnarray}
-2R^{t}{}_{t} = R^{\theta}{}_{\theta} &=& R^{\phi}{}_{\phi} = \ \frac{2a^{2}m}{\left(r^{2}+a^{2}\right)^{\frac{5}{2}}};
\qquad
R^{r}{}_{r} = \frac{a^{2}\left(3m-2\sqrt{r^{2}+a^{2}}\right)}{\left(r^{2}+a^{2}\right)^{\frac{5}{2}}}.
\end{eqnarray}
The Einstein tensor has non-zero (mixed) components:
\begin{eqnarray}
G^{t}{}_{t} &=& \frac{a^{2}\left(\sqrt{r^{2}+a^{2}}-4m\right)}{\left(r^{2}+a^{2}\right)^{\frac{5}{2}}} 
\qquad
G^{r}{}_{r} = \frac{-a^{2}}{\left(r^{2}+a^{2}\right)^{2}}\; \nonumber \\
G^{\theta}{}_{\theta} &=& G^{\phi}{}_{\phi} = \frac{a^{2}\left(\sqrt{r^{2}+a^{2}}-m\right)}{\left(r^{2}+a^{2}\right)^{\frac{5}{2}}} \ . 
\end{eqnarray}

%====================================================
\section{Curvature invariants}\label{sec:invariants}
%====================================================

The Ricci scalar is:
\begin{equation}
    R=\frac{2a^{2}\left(3m-\sqrt{r^{2}+a^{2}}\right)}{\left(r^{2}+a^{2}\right)^{\frac{5}{2}}} \ .
\end{equation}
The Ricci contraction $R_{ab}R^{ab}$ is:
\begin{equation}\label{3.10}
    R_{ab}R^{ab} = \frac{a^4\left[4\left(\sqrt{r^2+a^2}-\frac{3}{2}m\right)^{2}+(3m)^2\right]}{\left(r^2+a^2\right)^{5}} \ .
\end{equation}
Note that this is a sum of squares and so automatically non-negative (and finite).

The Weyl contraction $C_{abcd}C^{abcd}$:
\begin{equation}
    C_{abcd}C^{abcd} = \frac{4}{3\left(r^2+a^2\right)^{5}}\Bigg\lbrace 3m\left(2r^2-a^2\right)+2a^2\sqrt{r^2+a^2}\Bigg\rbrace^{2} \ .
\end{equation}
That this is a perfect square and so is automatically non-negative (and finite).

The Kretschmann scalar is:
\begin{equation}
R_{abcd} \, R^{abcd} = C_{abcd} \, C^{abcd} +2 R_{ab}\, R^{ab} - \frac{1}{3}R^2,
\end{equation}
and so (in view of the above) is guaranteed finite without further calculation.
Explicitly
\begin{eqnarray}
R_{abcd}\,R^{abcd} &=& \frac{4}{\left(r^{2}+a^{2}\right)^{5}}\bigg\lbrace\sqrt{r^{2}+a^{2}}\left[8a^2m\left(r^2-a^2\right)\right] \nonumber \\
&& +3a^{4}\left(r^2+a^2\right)+3m^{2}\left(3a^4-4a^2r^2+4r^4\right)\bigg\rbrace \ .
\end{eqnarray}

%===========================================
\section{Stress-energy tensor and energy conditions}
%===========================================

Let us examine the Einstein field equations for this spacetime. 
We first note that for $\sqrt{r^2+a^2}>2m$, that is, \emph{outside} any horizon that may potentially be present, one has $\rho = - T_t{}^t$ while $p_\parallel = T_r{}^r$ and $p_\perp =  T_\theta{}^\theta = T_\phi{}^\phi$. Using the mixed components $G^{\mu}_{\ \nu}=8\pi G_{N}\; T^{\mu}_{\ \nu}$, this
yields the following form of the stress-energy-momentum tensor:
\begin{eqnarray}
\rho &=& -\frac{a^{2}\left(\sqrt{r^{2}+a^{2}}-4m\right)}{8\pi G_{N}\left(r^{2}+a^{2}\right)^{\frac{5}{2}}} \ ; \nonumber \\
p_\parallel &=& \frac{-a^{2}}{8\pi G_{N}\left(r^{2}+a^{2}\right)^{2}} \ ; \nonumber \\
p_{\perp} &=& \frac{a^{2}\left(\sqrt{r^{2}+a^{2}}-m\right)}{8 \pi G_{N}\left(r^{2}+a^{2}\right)^{\frac{5}{2}}} \ .
\end{eqnarray}
Now a necessary condition for the NEC (null energy condition) to hold is that both $\rho + p_\parallel\geq 0$ and $\rho+p_\perp\geq 0$  for all $r$, $a$, $m$.  
It is sufficient to consider
\begin{eqnarray}
        \rho+p_\parallel  &=& \quad \frac{1}{8\pi G_{N}}
        \left\lbrace - \frac{a^2\left(\sqrt{r^2+a^2}-4m\right)}{\left(r^2+a^2\right)^{\frac{5}{2}}}-\frac{a^2}{\left(r^2+a^2\right)^{2}}\right\rbrace 
        \nonumber \\
        &=&  \frac{-a^2 (\sqrt{r^2+a^2}-2m) }{4\pi G_{N}\left(r^2+a^2\right)^{\frac{5}{2}}}.
\end{eqnarray}
Assuming $\sqrt{r^2+a^2}>2m$, this is manifestly negative for all values of $a$ and $m$ in our domain, and the NEC is clearly violated.

Note that for $\sqrt{r^2+a^2}<2m$, that is, \emph{inside} any horizon that may potentially be present, the $t$ and $r$ coordinates swap their timelike/spacelike characters and one has $\rho = - T_r{}^r$ while $p_\parallel = T_t{}^t$ and $p_\perp =  T_\theta{}^\theta = T_\phi{}^\phi$.
So inside the horizon
\begin{eqnarray}
\rho &=& \frac{a^{2}}{8\pi G_{N}\left(r^{2}+a^{2}\right)^{2}} \ ; \nonumber \\
p_\parallel &=&  \frac{a^{2}\left(\sqrt{r^{2}+a^{2}}-4m\right)}{8\pi G_{N}\left(r^{2}+a^{2}\right)^{\frac{5}{2}}},
\end{eqnarray}
and 
\begin{eqnarray}
        \rho+p_\parallel &=&  \frac{a^2 (\sqrt{r^2+a^2}-2m) }{4\pi G_{N}\left(r^2+a^2\right)^{\frac{5}{2}}}.
\end{eqnarray}
But since we are now working in the region $\sqrt{r^2+a^2}<2m$ this is again negative, and the NEC is again violated.

We can summarize this by stating
\begin{eqnarray}
        \rho+p_\parallel &=&  - \frac{a^2\; |\sqrt{r^2+a^2}-2m| }{4\pi G_{N}\left(r^2+a^2\right)^{\frac{5}{2}}},
\end{eqnarray}
which now holds for all values of $r$ and is negative everywhere except \emph{on} any horizon that may potentially be present. 

Demonstrating that the NEC is violated is sufficient to conclude that the weak, strong, and dominant energy conditions shall also be violated \cite{Lorentzian}. We therefore have a spacetime geometry that accurately models that of a regular black hole or a traversable wormhole depending on the value of $a$, but clearly violates all the classical energy conditions associated with the stress-energy-momentum tensor~\cite{Kar:2004, Molina-Paris:1998, Visser:cosmo1999, Barcelo:2000b, Visser:1999-super, Visser:1998-super, Abreu:2008, Abreu:2010, LNP, Martin-Moruno:2013a, Martin-Moruno:2013b,Martin-Moruno:2015, Visser:1994, Visser:1996a, Visser:1996b, Visser:1997-ec} .

%==============================
\section{Surface gravity and Hawking temperature}\label{sec:3+1}
%==============================

Let's calculate the surface gravity at the event horizon for the regular black hole case when $a\in(0,2m]$. The Killing vector which is null at the event horizon is $\xi^{\mu}=\partial_{t}$. This yields the following norm:
\begin{equation}
    \xi^{\mu}\xi_{\mu} = g_{\mu\nu}\xi^{\mu}\xi^{\nu} = g_{tt} = -\left(1-\frac{2m}{\sqrt{r^2+a^2}}\right) \ .
\end{equation}

Then we have the following relation for the surface gravity $\kappa$ (see for instance~\cite{Wald, MTW, Hell}):
\begin{equation}
    \nabla_{\nu}\left(-\xi^{\mu}\xi_{\mu}\right) = 2\kappa\xi_{\nu} \ .
\end{equation}
That is:
\begin{equation}
 \nabla_{\nu}\left(1-\frac{2m}{\sqrt{r^2+a^2}}\right) = 2\kappa\xi_{\nu} \ ;
\end{equation}
Keeping in mind that the event horizon is located at radial coordinate $r=\sqrt{(2m)^2-a^2}$ we see:
\begin{equation}
 \kappa = \frac{1}{2}\partial_{r}\left(1-\frac{2m}{\sqrt{r^2+a^2}}\right)\Bigg\vert_{r=\sqrt{(2m)^2-a^2}} = \frac{\sqrt{(2m)^2-a^2}}{8m^2} 
 = \kappa_\mathrm{Sch}\; \sqrt{1- {a^2\over(2m)^2}} .
\end{equation}
As a consistency check it is easily observed that for the Schwarzschild case when $a=0$, we have $\kappa=\frac{1}{4m}$, which is the expected surface gravity for the Schwarzschild black hole. For $a=2m$ the null horizon (one-way throat) is seen to be extremal.  It now follows that the temperature of Hawking radiation for our regular black hole is as follows (see for instance~\cite{Wald, MTW, Hell}):
\begin{eqnarray}
&& T_{H} = \frac{\hslash\kappa}{2\pi k_{B}}  = \frac{\hslash\sqrt{(2m)^2-a^2}}{16\pi k_{B}m^2} = T_{H,\mathrm{Sch}} \; \sqrt{1- {a^2\over(2m)^2}} .
\end{eqnarray}

%==============================
\section{ISCO and photon sphere analysis}\label{sec:isco}
%==============================

Let us now find the location of both the photon sphere for massless particles~\cite{Virbhadra:1999, Virbhadra:2002, Virbhadra:1998, Virbhadra:2007, Claudel:2000, Virbhadra:2008}  and the ISCO for massive particles as functions of $m$ and $a$.
Consider the tangent vector to the worldline of a massive or massless particle, parameterized by some arbitrary affine parameter, $\lambda$:
\begin{equation}
    g_{ab}\frac{dx^{a}}{d\lambda}\frac{dx^{b}}{d\lambda}=-g_{tt}\left(\frac{dt}{d\lambda}\right)^{2}+g_{rr}\left(\frac{dr}{d\lambda}\right)^{2}+\left(r^{2}+a^{2}\right)\left\lbrace\left(\frac{d\theta}{d\lambda}\right)^{2}+\sin^{2}\theta \left(\frac{d\phi}{d\lambda}\right)^{2}\right\rbrace \ .
\end{equation}
We may, without loss of generality, separate the two physically interesting cases (timelike and null) by defining 
\begin{equation}
    \epsilon = \left\{
    \begin{array}{rl}
    -1 & \qquad\mbox{massive particle, \emph{i.e.} timelike worldline} \\
     0 & \qquad\mbox{massless particle, \emph{i.e.} null worldline} .
    \end{array}\right. 
\end{equation}
That is, $ds^{2}/d\lambda^2=\epsilon$. Due to the metric being spherically symmetric we may fix $\theta=\frac{\pi}{2}$ arbitrarily and view the reduced equatorial problem:
\begin{equation}
    g_{ab}\frac{dx^{a}}{d\lambda}\frac{dx^{b}}{d\lambda}=-g_{tt}\left(\frac{dt}{d\lambda}\right)^{2}+g_{rr}\left(\frac{dr}{d\lambda}\right)^{2}+\left(r^{2}+a^{2}\right)\left(\frac{d\phi}{d\lambda}\right)^{2}=\epsilon \ .
\end{equation}

\noindent The Killing symmetries yield the following expressions for the conserved energy $E$ and angular momentum $L$ per unit mass (see for instance~\cite{Wald,MTW,Hell}):
\begin{equation}
    \left(1-\frac{2m}{\sqrt{r^{2}+a^{2}}}\right)\left(\frac{dt}{d\lambda}\right)=E \ ; \qquad\quad \left(r^{2}+a^{2}\right)\left(\frac{d\phi}{d\lambda}\right)=L \ .
\end{equation}
Hence:
\begin{equation}
    \left(1-\frac{2m}{\sqrt{r^{2}+a^{2}}}\right)^{-1}\left\lbrace -E^{2}+\left(\frac{dr}{d\lambda}\right)^{2}\right\rbrace+\frac{L^{2}}{r^{2}+a^{2}}=\epsilon \ ;
\end{equation}
implying
\begin{equation}
\left(\frac{dr}{d\lambda}\right)^{2}=E^{2}+\left(1-\frac{2m}{\sqrt{r^{2}+a^{2}}}\right)\left\lbrace\epsilon-\frac{L^{2}}{r^{2}+a^{2}}\right\rbrace \ .
\end{equation}
This gives ``effective potentials" for geodesic orbits as follows:
\begin{equation}
    V_{\epsilon}(r)=\left(1-\frac{2m}{\sqrt{r^{2}+a^{2}}}\right)\left\lbrace -\epsilon+\frac{L^{2}}{r^{2}+a^{2}}\right\rbrace \ .
\end{equation}
\begin{itemize}
    \item For a photon orbit we have the massless particle case $\epsilon=0$. Since we are in a spherically symmetric environment, solving for the locations of such orbits amounts to finding the coordinate location of the ``photon sphere''. That is, the value of the $r$-coordinate sufficiently close to our central mass such that photons are forced to propagate alomng circular geodesic orbits. These circular orbits occur at $V_{0}^{'}(r)=0$.  That is
    \begin{equation}
        V_{0}(r)=\left(1-\frac{2m}{\sqrt{r^{2}+a^{2}}}\right)\left(\frac{L^{2}}{r^{2}+a^{2}}\right) \ ,
    \end{equation}
leading to:
    \begin{equation}
        V_{0}^{'}(r)=\frac{2rL^{2}}{\left(r^{2}+a^{2}\right)^{\frac{5}{2}}}\bigg\lbrace 3m-\sqrt{r^{2}+a^{2}}\bigg\rbrace \ .
    \end{equation}
    When $V_{0}^{'}(r)=0$, if we discount the solution $r=0$ (as this spherical surface is clearly invalid for the location of the photon sphere), this gives the location of these circular orbits as $r=\pm\sqrt{(3m)^{2}-a^{2}}$. Firstly note that if $a\in(0,2m]$, $(3m)^{2}>a^{2}\ \forall \ a$, hence this solution does in fact correspond to a real-valued $r$-coordinate within our domain. Hence the photon sphere in our universe (\emph{i.e.} taking positive solution) for the case when the geometry is a regular black hole has coordinate location $r=\sqrt{(3m)^{2}-a^{2}}$. It also follows that in the case when $a>2m$ and we have a traversable wormhole, since we have strictly defined our $r$-coordinate to take on real values, there exists a photon sphere location in our universe only for the case when $2m<a<3m$. When $a>3m$ we have no photon sphere.
To verify stability, check the sign of $V_{0}^{''}(r)$:
    \begin{equation}
        V_{0}^{''}(r)=\frac{2L^{2}}{\left(r^{2}+a^{2}\right)^{\frac{7}{2}}}\Bigg\lbrace \sqrt{r^{2}+a^{2}}\left(3r^2-a^2\right)-3m\left(4r^2-a^2\right)\Bigg\rbrace \ .
    \end{equation}
For ease of notation let us first establish that when $r=\sqrt{(3m)^{2}-a^{2}}$, then $r^{2}+a^{2}=(3m)^{2}$, hence it can be shown that:
    \begin{equation}
        V_{0}^{''}\left(r=\sqrt{(3m)^{2}-a^{2}}\right)=\frac{-2L^{2}}{(3m)^{6}} \left((3m)^{2}-a^{2}\right)<0 \ .
    \end{equation}
Now $V^{''}_{0}<0$ implies instability, hence there is an unstable photon sphere at $r=\sqrt{(3m)^{2}-a^{2}}$ as presumed. For the Schwarzschild solution the location of the unstable photon sphere is at $r=3m$; which provides a useful consistency check.
      
    \item For massive particles the geodesic orbit corresponds to a timelike worldline and we have the case that $\epsilon=-1$. Therefore:
    \begin{equation}
        V_{-1}(r)=\left(1-\frac{2m}{\sqrt{r^{2}+a^{2}}}\right)\left(1+\frac{L^{2}}{r^{2}+a^{2}}\right) \ ,
    \end{equation}
    and it is easily verified that this leads to:
    \begin{equation}
        V_{-1}^{'}(r)=\frac{2r}{\left(r^{2}+a^{2}\right)^{\frac{5}{2}}}\bigg\lbrace L^{2}\left(3m-\sqrt{r^{2}+a^{2}}\right)+m\left(r^{2}+a^{2}\right)\bigg\rbrace \ .
    \end{equation}
    Equating this to zero and rearranging for $r$ gives a messy solution for $r$ as a function of $L$, $m$ and $a$. Instead it is preferable to assume a fixed circular orbit at some $r=r_{c}$, and rearrange the required angular momentum $L_{c}$ to be a function of $r_{c}$, $m$, and $a$. It then follows that the innermost circular orbit shall be the value of $r_{c}$ for which $L_{c}$ is minimised. % (since $L_{c}\sim r_{c}$ \red {(ref)}). 
    Hence if $V_{-1}^{'}(r_{c})=0$, we have:  
    \begin{equation}
        L_{c}^{2}\left(3m-\sqrt{r_{c}^{2}+a^{2}}\right)+m\left(r_{c}^{2}+a^{2}\right)=0 \ ,
    \end{equation}
implying    
    \begin{equation}
         L_{c}\left(r_{c}, m, a\right)=\sqrt{\frac{m\left(r_{c}^{2}+a^{2}\right)}{\sqrt{r_{c}^{2}+a^{2}}-3m}} \ ,
    \end{equation}
As a consistency check, for large $r_{c}$ (\emph{i.e.} $r_{c}>>a$) we observe that $L_{c}\approx\sqrt{mr_{c}}$, which is consistent with the expected value when considering circular orbits in weak-field GR.  Note that in classical physics the angular momentum per unit mass for a particle with angular velocity $\omega$ is $L_{c}\sim\omega r_{c}$. Kepler's third law of planetary motion implies that $\omega^2\sim {G_{N}m}/{r_{c}}$. (Here $m$ is the mass of the central object, as above.) It therefore follows that $L_{c}\sim\sqrt{{G_Nm}/{r_{c}}}\; r_{c}$. That is $L_{c}\sim\sqrt{mr_{c}}$, as above.
    
    \noindent It is then easily obtained that:
    \begin{equation}
        \frac{\partial L_{c}}{\partial r_{c}}=\left(\frac{\sqrt{m}r_{c}}{2\sqrt{\sqrt{r_{c}^{2}+a^{2}}-3m}}\right)\left(\frac{2}{\sqrt{r_{c}^{2}+a^{2}}}-\frac{1}{\sqrt{r_{c}^{2}+a^{2}}-3m}\right) \ .
    \end{equation}
    Solving for stationary points, and excluding $r_{c}=0$ (as this lies within the photon sphere, which is clearly an invalid solution for the ISCO of a massive particle):
    \begin{equation}
        \sqrt{r_{c}^{2}+a^{2}}-6m=0 \ ; \qquad \Longrightarrow\quad r_{c}=\sqrt{(6m)^{2}-a^{2}} \ ,
    \end{equation}
    (once again, discounting the negative solution for $r_{c}$ in the interests of remaining in our own universe). We therefore have a coordinate ISCO location at $r_{c}=\sqrt{(6m)^{2}-a^{2}}$. This is consistent with the expected value ($r=6m$) for Schwarzschild, when $a=0$. For our traversable wormhole geometry, provided $2m<a<6m$ we will have a valid ISCO location in our coordinate domain. When $a>6m$, we have a traversable wormhole with no ISCO.
\end{itemize}

\enlargethispage{10pt}
Denoting $r_{H}$ as the location of the horizon, $r_{\scriptscriptstyle{Photon}}$ as the location of the photon sphere, and $r_{\scriptscriptstyle{ISCO}}$ as the location of the ISCO, we have the following summary:
\begin{itemize}
    \item $r_{\scriptscriptstyle{H}}=\sqrt{(2m)^{2}-a^2}$ ;
    \item $r_{\scriptscriptstyle{Photon}}=\sqrt{(3m)^{2}-a^2}$ ;
    \item $r_{\scriptscriptstyle{ISCO}}=\sqrt{(6m)^{2}-a^2}$ .
\end{itemize}

%==============================
\section{Regge--Wheeler analysis}\label{sec:regge-wheeler}
%==============================

Considering the Regge-Wheeler Equation in view of the formalism developed in~\cite{Regge}, (see also reference~\cite{Boonserm:2018}), we may explicitly evaluate the Regge-Wheeler potentials for particles of spin $S\in\lbrace 0,1\rbrace$ in our spacetime. 
Firstly define a tortoise coordinate as follows 
\begin{equation}
dr_{*} = \frac{dr}{\left(1-\frac{2m}{\sqrt{r^2+a^2}}\right)} \ ,
\end{equation}
which gives the following expression for the metric (\ref{RBHmetric}):
\begin{equation}
    ds^2 = \left(1-\frac{2m}{\sqrt{r^2+a^2}}\right)\bigg\lbrace -dt^2+dr_{*}^2\bigg\rbrace+\left(r^2+a^2\right)\left(d\theta^2+\sin^2\theta\;\ d\phi^2\right) \ .
\end{equation}
It is convenient to write this as
\begin{equation}
    ds^2 = A(r_*)^2\bigg\lbrace -dt^2+dr_{*}^2\bigg\rbrace+B(r_*)^2\left(d\theta^2+\sin^2\theta \;d\phi^2\right) \ .
\end{equation}
The Regge--Wheeler equation is \cite{Regge}:
\begin{equation}
    \partial_{r_{*}}^{2}\hat{\phi}+\lbrace \omega^2-\mathcal{V}_S\rbrace\hat\phi = 0 \ ,
\end{equation}
where $\hat\phi$ is the scalar or vector field, $\mathcal{V}$ is the spin-dependent Regge-Wheeler potential for our particle, and $\omega$ is some temporal frequency component in the Fourier domain.
For a scalar field ($S=0$) examination of the d'Alembertian equation quickly yields
\begin{equation}
\mathcal{V}_{S=0} =   \left\lbrace{A^2 \over B^2} \right\rbrace \ell(\ell+1)
+ {\partial_{r_{*}}^2 B \over B} \ .
\end{equation}
For a vector field ($S=1$) conformal invariance in 3+1 dimensions guarantees that the Regge--Wheeler potential can depend only on the ratio $A/B$, whence normalizing to known results implies
\begin{equation}
\mathcal{V}_{S=1} =   \left\lbrace{A^2 \over B^2} \right\rbrace \ell(\ell+1).
\end{equation}
Collecting results, for $S\in\{0,1\}$ we have
\begin{equation}
\mathcal{V}_{S} =   \left\lbrace{A^2 \over B^2} \right\rbrace \ell(\ell+1)
+ (1-S) {\partial_{r_{*}}^2 B \over B} \ .
\end{equation}
The spin 2 axial mode is somewhat messier, and not of immediate interest. 

Noting that for our metric $\partial_{r_{*}}=\left(1-\frac{2m}{\sqrt{r^2+a^2}}\right)\partial_{r}$ and $B=\sqrt{r^2+a^2}$  we have:
\begin{equation}
    \frac{\partial_{r_{*}}^2 B}{B}=\left\lbrace 1-\frac{2m}{\sqrt{r^2+a^2}}\right\rbrace \; \left(2m(r^2-a^2) +a^2\sqrt{r^2+a^2} \over (r^2+a^2)^{5/2}\right).
\end{equation}
Therefore:
\begin{equation}
\mathcal{V}_{S\in\{0,1\}} =  \left(1-\frac{2m}{\sqrt{r^2+a^2}}\right) \left\lbrace{\ell(\ell+1)\over r^2+a^2} 
+ (1-S) \left(2m(r^2-a^2) +a^2\sqrt{r^2+a^2} \over (r^2+a^2)^{5/2}\right) \right\rbrace.
\end{equation}
This has the correct behaviour as $a\to0$. Note that this Regge--Wheeler potential is symmetric about $r=0$. 
For $a<2m$ the situation is qualitatively similar to the usual Schwarzschild case (the tortoise coordinate diverges at either horizon, and $\mathcal{V}_{S\in\{0,1\}} \to 0$ at either horizon).  For $a=2m$ the tortoise coordinate diverges at the extremal horizon (one-way null throat), while we still have $\mathcal{V}_{S\in\{0,1\}} \to 0$. For $a>2m$ the tortoise coordinate converges at the wormhole throat, while we now have have $\mathcal{V}_{S\in\{0,1\}}$ nonzero and positive at the throat:
\begin{equation}
\mathcal{V}_{S\in\{0,1\}} \to  \left(1-\frac{2m}{a}\right) \left\lbrace{\ell(\ell+1)\over a^2} 
+ (1-S) \left(a-2m \over a^3\right) \right\rbrace.
\end{equation}

%========================================================
\section{Discussion}\label{sec:dis}
%========================================================

The regular black hole presented above in some sense represents minimal violence to the standard Schwarzschild solution. 
Indeed for $a=0$ it is the standard Schwarzschild solution.  For $a\in(0,2m)$ the Carter--Penrose diagram is in some sense ``as close as possible'' to that for the maximally extended Kruskal--Szekeres version of Schwarzschild, except that the singularity is converted into a spacelike hypersurface representing a ``bounce'' into a future incarnation of the universe. This is qualitatively different from the picture where the collapsing regular black hole ``bounces'' back into our own universe~\cite{Barcelo:2014, Barcelo:2014b, Barcelo:2015, Barcelo:2016, Garay:2017, Rovelli:2014, Haggard:2015, Christodoulou:2016, DeLorenzo:2015, Malafarina:2017, Olmedo:2017, Barrau:2018, Malafarina:2018},  and is a scenario that deserves some attention in its own right. The specific model introduced above also has the very nice feature that it analytically interpolates between black holes and traversable wormholes in a particularly clear and tractable manner. 
The ``one-way'' wormhole at $a=2m$, where the throat becomes null and extremal, is particularly interesting and novel.

%========================================================
\section*{Acknowledgments}
%========================================================
MV acknowledges financial support via the Marsden Fund administered by the Royal Society of New Zealand.
MV wishes to thank Sayan Kar for useful comments and questions.

%========================================================
%========================================================

%========================================================

\begin{thebibliography}{69}  
%========================================================
%========================================================

%========================================================
% REGULAR BLACK HOLES
%========================================================

\bibitem{Bardeen:1968}
J.~M.~Bardeen, ``Non-singular general-relativistic gravitational collapse'', 
in Proceedings of International Conference GR5, 1968, Tbilisi, USSR, p. 174.

\bibitem{Bergmann-Roman}
Thomas A. Roman and Peter G. Bergmann, 
``Stellar collapse without singularities?'',
Phys. Rev. D {\bf28} (1983) 1265--1277.\\
doi: \url{https://doi.org/10.1103/PhysRevD.28.1265}

\bibitem{Hayward:2005}
  S.~A.~Hayward,
  ``Formation and evaporation of regular black holes'',\\
  Phys.\ Rev.\ Lett.\  {\bf 96} (2006) 031103
  doi:10.1103/PhysRevLett.96.031103
  [gr-qc/0506126].
  %%CITATION = doi:10.1103/PhysRevLett.96.031103;%%
  %309 citations counted in INSPIRE as of 30 Oct 2018
  
\bibitem{Bardeen:2014}
  J.~M.~Bardeen,
  ``Black hole evaporation without an event horizon'',\\
  arXiv:1406.4098 [gr-qc].
  %%CITATION = ARXIV:1406.4098;%%
  %40 citations counted in INSPIRE as of 30 Oct 2018
  
  \bibitem{Frolov:2014} 
  V.~P.~Frolov,\\
  ``Information loss problem and a black hole model with a closed apparent horizon'',\\
  JHEP {\bf 1405}, 049 (2014)
  doi:10.1007/JHEP05(2014)049
  [arXiv:1402.5446 [hep-th]].
  %%CITATION = doi:10.1007/JHEP05(2014)049;%%
  %65 citations counted in INSPIRE as of 30 Oct 2018
  

 \bibitem{Frolov:2014b}
  V.~P.~Frolov,
  ``Do Black Holes Exist?'',
  arXiv:1411.6981 [hep-th].
  %%CITATION = ARXIV:1411.6981;%%
  %20 citations counted in INSPIRE as of 11 Dec 2018
 
 \bibitem{Frolov:2016}
  V.~P.~Frolov,
  ``Notes on nonsingular models of black holes'',
  Phys.\ Rev.\ D {\bf 94} (2016) no.10,  104056
  doi:10.1103/PhysRevD.94.104056
  [arXiv:1609.01758 [gr-qc]].
  %%CITATION = doi:10.1103/PhysRevD.94.104056;%%
  %44 citations counted in INSPIRE as of 11 Dec 2018
 
 \bibitem{Frolov:2017}
  V.~P.~Frolov and A.~Zelnikov,
  ``Quantum radiation from an evaporating nonsingular black hole'',
  Phys.\ Rev.\ D {\bf 95} (2017) no.12,  124028
  doi:10.1103/PhysRevD.95.124028
  [arXiv:1704.03043 [hep-th]].
  %%CITATION = doi:10.1103/PhysRevD.95.124028;%%
  %14 citations counted in INSPIRE as of 11 Dec 2018

  
  \bibitem{Frolov:2018}
  V.~P.~Frolov,
  ``Remarks on non-singular black holes'',
  EPJ Web Conf.\  {\bf 168} (2018) 01001
  doi:10.1051/epjconf/201816801001
  [arXiv:1708.04698 [gr-qc]].
  %%CITATION = doi:10.1051/epjconf/201816801001;%%
  
  \bibitem{Cano:2018}
  P.~A.~Cano, S.~Chimento, T.~Ort\'in and A.~Ruip\'erez,
  ``Regular Stringy Black Holes?,''
  arXiv:1806.08377 [hep-th].
  %%CITATION = ARXIV:1806.08377;%%
  %4 citations counted in INSPIRE as of 23 Dec 2018
  
  \bibitem{Bardeen:2018}
  J.~M.~Bardeen,
  ``Models for the nonsingular transition of an evaporating black hole into a white hole'',
  arXiv:1811.06683 [gr-qc].
  %%CITATION = ARXIV:1811.06683;%%
  
  \bibitem{regular}
  R.~Carballo-Rubio, F.~Di Filippo, S.~Liberati, C.~Pacilio and M.~Visser,
  ``On the viability of regular black holes'',
  JHEP {\bf 1807} (2018) 023
   [JHEP {\bf 2018} (2020) 023]
  doi:10.1007/JHEP07(2018)023
  [arXiv:1805.02675 [gr-qc]].
  %%CITATION = doi:10.1007/JHEP07(2018)023;%%
  %5 citations counted in INSPIRE as of 31 Oct 2018
  
 \bibitem{beyond}
  R.~Carballo-Rubio, F.~Di Filippo, S.~Liberati and M.~Visser,
  ``Phenomenological aspects of black holes beyond general relativity'',
  Physical Review D {\bf98} (2018) 124009.\\
  doi: 10.1103/PhysRevD.98.124009 [arXiv:1809.08238 [gr-qc]].
  %%CITATION = ARXIV:1809.08238;%%
  %2 citations counted in INSPIRE as of 31 Oct 2018 

%========================================================
% WORMHOLES
%========================================================

\bibitem{Morris:1988a}
  M.~S.~Morris and K.~S.~Thorne,
  ``Wormholes in space-time and their use for interstellar travel: A tool for teaching general relativity'',
  Am.\ J.\ Phys.\  {\bf 56} (1988) 395.
  doi:10.1119/1.15620
  %%CITATION = doi:10.1119/1.15620;%%
  %1139 citations counted in INSPIRE as of 30 Oct 2018
  

\bibitem{Morris:1988b}
  M.~S.~Morris, K.~S.~Thorne and U.~Yurtsever,
  ``Wormholes, Time Machines, and the Weak Energy Condition'',
  Phys.\ Rev.\ Lett.\  {\bf 61} (1988) 1446.
  doi:10.1103/PhysRevLett.61.1446
  %%CITATION = doi:10.1103/PhysRevLett.61.1446;%%
  %813 citations counted in INSPIRE as of 30 Oct 2018
  
  \bibitem{Visser:1989a}
  M.~Visser,
  ``Traversable wormholes: Some simple examples'',\\
  Phys.\ Rev.\ D {\bf 39} (1989) 3182
  doi:10.1103/PhysRevD.39.3182
  [arXiv:0809.0907 [gr-qc]].
  %%CITATION = doi:10.1103/PhysRevD.39.3182;%%
  %238 citations counted in INSPIRE as of 30 Oct 2018
  
%\cite{Visser:1989kg}
\bibitem{Visser:1989b}
  M.~Visser,
  ``Traversable wormholes from surgically modified Schwarzschild space-times'',
  Nucl.\ Phys.\ B {\bf 328} (1989) 203
  doi:10.1016/0550-3213(89)90100-4
  [arXiv:0809.0927 [gr-qc]].
  %%CITATION = doi:10.1016/0550-3213(89)90100-4;%%
  %208 citations counted in INSPIRE as of 30 Oct 2018
  
 \bibitem{Lorentzian}
  M.~Visser, ``Lorentzian wormholes: From Einstein to Hawking", \\
  (AIP Press, now Springer, New York, 1995).
  
\bibitem{Visser:2003}
  M.~Visser, S.~Kar and N.~Dadhich,
  ``Traversable wormholes with arbitrarily small energy condition violations'',
  Phys.\ Rev.\ Lett.\  {\bf 90} (2003) 201102
  doi:10.1103/PhysRevLett.90.201102
  [gr-qc/0301003].
  %%CITATION = doi:10.1103/PhysRevLett.90.201102;%%
  %200 citations counted in INSPIRE as of 30 Oct 2018
  

\bibitem{Hochberg:1997}
  D.~Hochberg and M.~Visser,\\
  ``Geometric structure of the generic static traversable wormhole throat'',\\
  Phys.\ Rev.\ D {\bf 56} (1997) 4745
  doi:10.1103/PhysRevD.56.4745
  [gr-qc/9704082].
  %%CITATION = doi:10.1103/PhysRevD.56.4745;%%
  %192 citations counted in INSPIRE as of 30 Oct 2018
  
\bibitem{Poisson:1995}
  E.~Poisson and M.~Visser,
  ``Thin shell wormholes: Linearization stability'',\\
  Phys.\ Rev.\ D {\bf 52} (1995) 7318
  doi:10.1103/PhysRevD.52.7318
  [gr-qc/9506083].
  %%CITATION = doi:10.1103/PhysRevD.52.7318;%%
  %192 citations counted in INSPIRE as of 30 Oct 2018
  

\bibitem{Barcelo:2000}
  C.~Barcel\'o and M.~Visser,
  ``Scalar fields, energy conditions, and traversable wormholes'',
  Class.\ Quant.\ Grav.\  {\bf 17} (2000) 3843
  doi:10.1088/0264-9381/17/18/318
  [gr-qc/0003025].
  %%CITATION = doi:10.1088/0264-9381/17/18/318;%%
  %129 citations counted in INSPIRE as of 30 Oct 2018
  

\bibitem{Hochberg:1998}
  D.~Hochberg and M.~Visser,
  ``The Null energy condition in dynamic wormholes'',
  Phys.\ Rev.\ Lett.\  {\bf 81} (1998) 746
  doi:10.1103/PhysRevLett.81.746
  [gr-qc/9802048].
  %%CITATION = doi:10.1103/PhysRevLett.81.746;%%
  %123 citations counted in INSPIRE as of 30 Oct 2018
  
\bibitem{Cramer:1994}
  J.~G.~Cramer, R.~L.~Forward, M.~S.~Morris, M.~Visser, G.~Benford and G.~A.~Landis,
  ``Natural wormholes as gravitational lenses'',
  Phys.\ Rev.\ D {\bf 51} (1995) 3117
  doi:10.1103/PhysRevD.51.3117
  [astro-ph/9409051].
  %%CITATION = doi:10.1103/PhysRevD.51.3117;%%
  %122 citations counted in INSPIRE as of 30 Oct 2018
  
 
\bibitem{Visser:1997}
  M.~Visser and D.~Hochberg,
  ``Generic wormhole throats'',
  Annals Israel Phys.\ Soc.\  {\bf 13} (1997) 249
  [gr-qc/9710001].
  %%CITATION = GR-QC/9710001;%%
  %41 citations counted in INSPIRE as of 30 Oct 2018


 
\bibitem{Barcelo:1999}
  C.~Barcel\'o and M.~Visser,
  ``Traversable wormholes from massless conformally coupled scalar fields'',
  Phys.\ Lett.\ B {\bf 466} (1999) 127
  doi:10.1016/S0370-2693(99)01117-X
  [gr-qc/9908029].
  %%CITATION = doi:10.1016/S0370-2693(99)01117-X;%%
  %113 citations counted in INSPIRE as of 30 Oct 2018
  
    
\bibitem{Garcia:2011}
  N.~M.~Garcia, F.~S.~N.~Lobo and M.~Visser,
  ``Generic spherically symmetric dynamic thin-shell traversable wormholes in standard general relativity'',
  Phys.\ Rev.\ D {\bf 86} (2012) 044026
  doi:10.1103/PhysRevD.86.044026
  [arXiv:1112.2057 [gr-qc]].
  %%CITATION = doi:10.1103/PhysRevD.86.044026;%%
  %75 citations counted in INSPIRE as of 30 Oct 2018
  
\bibitem{Boonserm:2018}
  P.~Boonserm, T.~Ngampitipan, A.~Simpson and M.~Visser,
  ``The exponential metric represents a traversable wormhole'',
  arXiv:1805.03781 [gr-qc].
  %%CITATION = ARXIV:1805.03781;%%
  %2 citations counted in INSPIRE as of 30 Oct 2018
  
  \bibitem{Lobo:2004}
  F.~S.~N.~Lobo,
  ``Thin shells around traversable wormholes'',
  gr-qc/0401083.
  %%CITATION = GR-QC/0401083;%%
  %1 citations counted in INSPIRE as of 30 Oct 2018

  %==============================================================
  % BOUNCE --- into our own universe
  %==============================================================
  
  
  \bibitem{Barcelo:2014}
  C.~Barcel\'o, R.~Carballo-Rubio, L.~J.~Garay and G.~Jannes,
  ``The lifetime problem of evaporating black holes: mutiny or resignation,''
  Class.\ Quant.\ Grav.\  {\bf 32} (2015) no.3,  035012
  doi:10.1088/0264-9381/32/3/035012
  [arXiv:1409.1501 [gr-qc]].
  %%CITATION = doi:10.1088/0264-9381/32/3/035012;%%
  %38 citations counted in INSPIRE as of 17 Dec 2018
  
 \bibitem{Barcelo:2014b}
  C.~Barcel\'o, R.~Carballo-Rubio and L.~J.~Garay,
  ``Mutiny at the white-hole district,''
  Int.\ J.\ Mod.\ Phys.\ D {\bf 23} (2014) no.12,  1442022
  doi:10.1142/S021827181442022X
  [arXiv:1407.1391 [gr-qc]].
  %%CITATION = doi:10.1142/S021827181442022X;%%
  %20 citations counted in INSPIRE as of 17 Dec 2018 
  
  \bibitem{Barcelo:2015}
  C.~Barcel\'o, R.~Carballo-Rubio, L.~J.~Garay and G.~Jannes,
  ``Do transient white holes have a place in Nature?,''
  J.\ Phys.\ Conf.\ Ser.\  {\bf 600} (2015) no.1,  012033.
  doi:10.1088/1742-6596/600/1/012033
  %%CITATION = doi:10.1088/1742-6596/600/1/012033;%%
  
  \bibitem{Barcelo:2016}
  C.~Barcel\'o, R.~Carballo-Rubio and L.~J.~Garay,
  ``Exponential fading to white of black holes in quantum gravity,''
  Class.\ Quant.\ Grav.\  {\bf 34} (2017) no.10,  105007
  doi:10.1088/1361-6382/aa6962
  [arXiv:1607.03480 [gr-qc]].
  %%CITATION = doi:10.1088/1361-6382/aa6962;%%
  %13 citations counted in INSPIRE as of 17 Dec 2018
  
  \bibitem{Garay:2017}
  L.~J.~Garay, C.~Barcel\'o, R.~Carballo-Rubio and G.~Jannes,
  ``Do stars die too long?,''
  doi:10.1142/9789813226609\_0174
  %%CITATION = doi:10.1142/9789813226609_0174;%%
  %2 citations counted in INSPIRE as of 17 Dec 2018
  
  \bibitem{Rovelli:2014}
  C.~Rovelli and F.~Vidotto,
  ``Planck stars,''
  Int.\ J.\ Mod.\ Phys.\ D {\bf 23} (2014) no.12,  1442026
  doi:10.1142/S0218271814420267
  [arXiv:1401.6562 [gr-qc]].
  %%CITATION = doi:10.1142/S0218271814420267;%%
  %112 citations counted in INSPIRE as of 17 Dec 2018
  
  \bibitem{Haggard:2015}
  H.~M.~Haggard and C.~Rovelli,
  ``Black to white hole tunneling: An exact classical solution,''
  Int.\ J.\ Mod.\ Phys.\ A {\bf 30} (2015) no.28n29,  1545015.
  doi:10.1142/S0217751X15450153
  %%CITATION = doi:10.1142/S0217751X15450153;%%
  %4 citations counted in INSPIRE as of 17 Dec 2018
  
  \bibitem{Christodoulou:2016}
  M.~Christodoulou, C.~Rovelli, S.~Speziale and I.~Vilensky,
  ``Planck star tunneling time: An astrophysically relevant observable from background-free quantum gravity,''
  Phys.\ Rev.\ D {\bf 94} (2016) no.8,  084035
  doi:10.1103/PhysRevD.94.084035
  [arXiv:1605.05268 [gr-qc]].
  %%CITATION = doi:10.1103/PhysRevD.94.084035;%%
  %46 citations counted in INSPIRE as of 17 Dec 2018
  
  \bibitem{DeLorenzo:2015}
  T.~De Lorenzo and A.~Perez,
  ``Improved Black Hole Fireworks: Asymmetric Black-Hole-to-White-Hole Tunneling Scenario,''
  Phys.\ Rev.\ D {\bf 93} (2016) no.12,  124018
  doi:10.1103/PhysRevD.93.124018
  [arXiv:1512.04566 [gr-qc]].
  %%CITATION = doi:10.1103/PhysRevD.93.124018;%%
  %23 citations counted in INSPIRE as of 17 Dec 2018
  
  \bibitem{Malafarina:2017}
  D.~Malafarina,
  ``Classical collapse to black holes and quantum bounces: A review,''
  Universe {\bf 3} (2017) no.2,  48
  doi:10.3390/universe3020048
  [arXiv:1703.04138 [gr-qc]].
  %%CITATION = doi:10.3390/universe3020048;%%
  %13 citations counted in INSPIRE as of 17 Dec 2018

\bibitem{Olmedo:2017}
  J.~Olmedo, S.~Saini and P.~Singh,
  ``From black holes to white holes: a quantum gravitational, symmetric bounce,''
  Class.\ Quant.\ Grav.\  {\bf 34} (2017) no.22,  225011
  doi:10.1088/1361-6382/aa8da8
  [arXiv:1707.07333 [gr-qc]].
  %%CITATION = doi:10.1088/1361-6382/aa8da8;%%
  %17 citations counted in INSPIRE as of 17 Dec 2018
  
  \bibitem{Barrau:2018}
  A.~Barrau, K.~Martineau and F.~Moulin,
  ``A status report on the phenomenology of black holes in loop quantum gravity: Evaporation, tunneling to white holes, dark matter and gravitational waves,''
  Universe {\bf 4} (2018) no.10,  102
  doi:10.3390/universe4100102
  [arXiv:1808.08857 [gr-qc]].
  %%CITATION = doi:10.3390/universe4100102;%%
  %2 citations counted in INSPIRE as of 17 Dec 2018

  \bibitem{Malafarina:2018}
  D.~Malafarina,
  ``Black Hole Bounces on the Road to Quantum Gravity,''
  Universe {\bf 4} (2018) no.9,  92.
  doi:10.3390/universe4090092
  %%CITATION = doi:10.3390/universe4090092;%%
  %1 citations counted in INSPIRE as of 17 Dec 2018
  
 %==============================================================
 % GENERIC
 %==============================================================
   
\bibitem{Wald}
 R.~M.~Wald, \emph{Gravitation}, (University of Chicago Press, USA, 1984).
 
\bibitem{MTW}
  C.~W.~Misner, K.~S.~Thorne, and J.~A.~Wheeler, \emph{Gravitation}, \\
  (Freeman, San Francisco, 1973). 


\bibitem{Hell} S. W. Hawking and G. F. R. Ellis,
\emph{The large scale structure of space-time},\\
(Cambridge University Press, Cambridge, 1973).

 %==============================================================
 % TWISTED
 %============================================================== 
   \bibitem{twisted}
  F.~Gray, J.~Santiago, S.~Schuster and M.~Visser,
  ``Twisted black holes are unphysical'',
  Mod.\ Phys.\ Lett.\ A {\bf 32} (2017) no.18,  1771001
  doi:10.1142/S0217732317710018
  [arXiv:1610.06135 [gr-qc]].
  %%CITATION = doi:10.1142/S0217732317710018;%%
  %1 citations counted in INSPIRE as of 11 Dec 2018

  %==========================================================
  % ENERGY CONDITIONS
  %==========================================================

\bibitem{Kar:2004}
  S.~Kar, N.~Dadhich and M.~Visser,
  ``Quantifying energy condition violations in traversable wormholes'',
  Pramana {\bf 63} (2004) 859
  doi:10.1007/BF02705207
  [gr-qc/0405103].
  %%CITATION = doi:10.1007/BF02705207;%%
  %44 citations counted in INSPIRE as of 30 Oct 2018


\bibitem{Molina-Paris:1998}
  C.~Molina-Par\'is and M.~Visser,
  ``Minimal conditions for the creation of a Friedman-Robertson-Walker universe from a bounce'',
  Phys.\ Lett.\ B {\bf 455} (1999) 90
  doi:10.1016/S0370-2693(99)00469-4
  [gr-qc/9810023].
  %%CITATION = doi:10.1016/S0370-2693(99)00469-4;%%
  %81 citations counted in INSPIRE as of 30 Oct 2018
  
  \bibitem{Visser:cosmo1999}
  M.~Visser and C.~Barcel\'o,
  ``Energy conditions and their cosmological implications'',
  doi:10.1142/9789812792129\_0014
  gr-qc/0001099.
  %%CITATION = doi:10.1142/9789812792129_0014;%%
  %74 citations counted in INSPIRE as of 30 Oct 2018
  
  \bibitem{Barcelo:2000b}
  C.~Barcel\'o and M.~Visser,
  ``Brane surgery: Energy conditions, traversable wormholes, and voids'',
  Nucl.\ Phys.\ B {\bf 584} (2000) 415
  doi:10.1016/S0550-3213(00)00379-5
  [hep-th/0004022].
  %%CITATION = doi:10.1016/S0550-3213(00)00379-5;%%
  %68 citations counted in INSPIRE as of 30 Oct 2018
  
  \bibitem{Visser:1999-super}
  M.~Visser, B.~Bassett and S.~Liberati,
  ``Perturbative superluminal censorship and the null energy condition'',
  AIP Conf.\ Proc.\  {\bf 493} (1999) no.1,  301
  doi:10.1063/1.1301601
  [gr-qc/9908023].
  %%CITATION = doi:10.1063/1.1301601;%%
  %18 citations counted in INSPIRE as of 30 Oct 2018

%======================================================
  
\bibitem{Visser:1998-super}
  M.~Visser, B.~Bassett and S.~Liberati,
  ``Superluminal censorship'',
  Nucl.\ Phys.\ Proc.\ Suppl.\  {\bf 88} (2000) 267
  doi:10.1016/S0920-5632(00)00782-9
  [gr-qc/9810026].
  %%CITATION = doi:10.1016/S0920-5632(00)00782-9;%%
  %61 citations counted in INSPIRE as of 30 Oct 2018
    

\bibitem{Abreu:2008}
  G.~Abreu and M.~Visser,
  ``Quantum Interest in (3+1) dimensional Minkowski space'',
  Phys.\ Rev.\ D {\bf 79} (2009) 065004
  doi:10.1103/PhysRevD.79.065004
  [arXiv:0808.1931 [gr-qc]].
  %%CITATION = doi:10.1103/PhysRevD.79.065004;%%
  %6 citations counted in INSPIRE as of 30 Oct 2018
  



\bibitem{Abreu:2010}
  G.~Abreu and M.~Visser,
  ``The Quantum interest conjecture in (3+1)-dimensional Minkowski space'',
  doi:10.1142/9789814374552\_0481
  arXiv:1001.1180 [gr-qc].
  %%CITATION = doi:10.1142/9789814374552_0481;%%
  %2 citations counted in INSPIRE as of 30 Oct 2018

%========================================================
%========================================================
\bibitem{LNP}
  P.~Mart\'in--Moruno and M.~Visser,
  ``Classical and semi-classical energy conditions'',\\
  Fundam.\ Theor.\ Phys.\  {\bf 189} (2017) 193 (Lecture Notes in Physics)\\
  doi:10.1007/978-3-319-55182-1\_9
  [arXiv:1702.05915 [gr-qc]].
  %%CITATION = doi:10.1007/978-3-319-55182-1_9;%%
  %2 citations counted in INSPIRE as of 30 Jun 2017
  
 
%========================================================
% CLASSICAL ENERGY CONDITIONS (MORE)
%========================================================
%========================================================
% SEMI-CLASSICAL ENERGY CONDITIONS (MORE) 
%========================================================
  
 %========================================================

\bibitem{Martin-Moruno:2013a} 
  P.~Mart\'in--Moruno and M.~Visser,
  ``Classical and quantum flux energy conditions for quantum vacuum states'',
  Phys.\ Rev.\ D {\bf 88}, no. 6, 061701 (2013)
  doi:10.1103/PhysRevD.88.061701
  [arXiv:1305.1993 [gr-qc]].
  %%CITATION = doi:10.1103/PhysRevD.88.061701;%%
  %18 citations counted in INSPIRE as of 30 Jun 2017
  %========================================================

\bibitem{Martin-Moruno:2013b}
  P.~Mart\'in--Moruno and M.~Visser,
  ``Semiclassical energy conditions for quantum vacuum states'',
  JHEP {\bf 1309} (2013) 050
  doi:10.1007/JHEP09(2013)050
  [arXiv:1306.2076 [gr-qc]].
  %%CITATION = doi:10.1007/JHEP09(2013)050;%%
  %18 citations counted in INSPIRE as of 30 Jun 2017
%========================================================
  
  \bibitem{Martin-Moruno:2015}
  P.~Mart\'in--Moruno and M.~Visser,
  ``Semi-classical and nonlinear energy conditions'', 
  Proceedings of The Fourteenth Marcel Grossmann Meeting, 1442-1447. World Scientific 2017.
  doi:10.1142/9789813226609\_0126 
  [arXiv:1510.00158 [gr-qc]].
  %%CITATION = ARXIV:1510.00158;%%
  %3 citations counted in INSPIRE as of 30 Jun 2017
%========================================================

%========================================================
  
  \bibitem{Visser:1994}
  M.~Visser,
  ``Scale anomalies imply violation of the averaged null energy condition'',\\
  Phys.\ Lett.\ B {\bf 349} (1995) 443
  doi:10.1016/0370-2693(95)00303-3
  [gr-qc/9409043].
  %%CITATION = doi:10.1016/0370-2693(95)00303-3;%%
  %49 citations counted in INSPIRE as of 13 Jul 2017
 %========================================================
 %\enlargethispage{40pt}
 \bibitem{Visser:1996a}
  M.~Visser,
  ``Gravitational vacuum polarization. 1: Energy conditions in the Hartle-Hawking vacuum'',
  Phys.\ Rev.\ D {\bf 54} (1996) 5103
  doi:10.1103/PhysRevD.54.5103
  [gr-qc/9604007].
  %%CITATION = doi:10.1103/PhysRevD.54.5103;%%
  %58 citations counted in INSPIRE as of 30 Jan 2018
  
  \bibitem{Visser:1996b}
  M.~Visser,
  ``Gravitational vacuum polarization. 2: Energy conditions in the Boulware vacuum'',
  Phys.\ Rev.\ D {\bf 54} (1996) 5116
  doi:10.1103/PhysRevD.54.5116
  [gr-qc/9604008].
  %%CITATION = doi:10.1103/PhysRevD.54.5116;%%
  %68 citations counted in INSPIRE as of 30 Jan 2018
  
\bibitem{Visser:1997-ec}
  M.~Visser,
  ``Gravitational vacuum polarization. 4: Energy conditions in the Unruh vacuum'',
  Phys.\ Rev.\ D {\bf 56} (1997) 936
  doi:10.1103/PhysRevD.56.936
  [gr-qc/9703001].
  %%CITATION = doi:10.1103/PhysRevD.56.936;%%
  %49 citations counted in INSPIRE as of 13 Jul 2017
%========================================================
%========================================================
%========================================================
  %%% PHOTON SPHERE
  %========================================================
  
  \bibitem{Virbhadra:1999}
  K.~S.~Virbhadra and G.~F.~R.~Ellis,
  ``Schwarzschild black hole lensing'',
  Phys.\ Rev.\ D {\bf 62} (2000) 084003
  doi:10.1103/PhysRevD.62.084003
  [astro-ph/9904193].
  %%CITATION = doi:10.1103/PhysRevD.62.084003;%%
  %408 citations counted in INSPIRE as of 17 May 2018

\bibitem{Virbhadra:2002}
  K.~S.~Virbhadra and G.~F.~R.~Ellis,
  ``Gravitational lensing by naked singularities'',
  Phys.\ Rev.\ D {\bf 65} (2002) 103004.
  doi:10.1103/PhysRevD.65.103004
  %%CITATION = doi:10.1103/PhysRevD.65.103004;%%
  %327 citations counted in INSPIRE as of 17 May 2018

\bibitem{Virbhadra:1998}
  K.~S.~Virbhadra, D.~Narasimha and S.~M.~Chitre,
  ``Role of the scalar field in gravitational lensing'',
  Astron.\ Astrophys.\  {\bf 337} (1998) 1
  [astro-ph/9801174].
  %%CITATION = ASTRO-PH/9801174;%%
  %219 citations counted in INSPIRE as of 17 May 2018

\bibitem{Virbhadra:2007}
  K.~S.~Virbhadra and C.~R.~Keeton,
  ``Time delay and magnification centroid due to gravitational lensing by black holes and naked singularities'',
  Phys.\ Rev.\ D {\bf 77} (2008) 124014
  doi:10.1103/PhysRevD.77.124014
  [arXiv:0710.2333 [gr-qc]].
  %%CITATION = doi:10.1103/PhysRevD.77.124014;%%
  %205 citations counted in INSPIRE as of 17 May 2018

\bibitem{Claudel:2000}
  C.~M.~Claudel, K.~S.~Virbhadra and G.~F.~R.~Ellis,
  ``The Geometry of photon surfaces'',
  J.\ Math.\ Phys.\  {\bf 42} (2001) 818
  doi:10.1063/1.1308507
  [gr-qc/0005050].
  %%CITATION = doi:10.1063/1.1308507;%%
  %184 citations counted in INSPIRE as of 17 May 2018

\bibitem{Virbhadra:2008}
  K.~S.~Virbhadra,
  ``Relativistic images of Schwarzschild black hole lensing'',
  Phys.\ Rev.\ D {\bf 79} (2009) 083004
  doi:10.1103/PhysRevD.79.083004
  [arXiv:0810.2109 [gr-qc]].
  %%CITATION = doi:10.1103/PhysRevD.79.083004;%%
  %169 citations counted in INSPIRE as of 17 May 2018
  
  %========================================================

   
 
%==============================================================
% REGGE-WHEELER
%==============================================================
 \bibitem{Regge}
P.~Boonserm, T.~Ngampitipan and M.~Visser,
  ``Regge--Wheeler equation, linear stability, and greybody factors for dirty black holes'',\\
  Phys.\ Rev.\ D {\bf 88} (2013) 041502
  doi:10.1103/PhysRevD.88.041502 \\{}
  [arXiv:1305.1416 [gr-qc]].
  %%CITATION = doi:10.1103/PhysRevD.88.041502;%%
  %12 citations counted in INSPIRE as of 08 Apr 2018
 %==============================================================
 

  
  
 %========================================================
\end{thebibliography}
\end{document}